\documentstyle[preprint,aps]{revtex}

\begin{document}
\title{Experimental Implementation of Remote State Preparation by Nuclear Magnetic
Resonance}
\author{Xinhua Peng$^{1}$, Xiwen Zhu$^{1\thanks{%
Corresponding author. {E-mail:xwzhu@wipm.whcnc.ac.cn; Fax: 0086-27-87885291.}%
}}$, Ximing Fang$^{2,1}$, Mang Feng$^{1}$, Maili Liu$^{1},$ and Kelin Gao$%
^{1}$}
\address{$^{1}$Laboratory of Magnetic Resonance and Molecular Physics, \\
Wuhan Institute of Physics and Mathematics, the Chinese Academy of Sciences, 
\\
Wuhan, 430071, People's Republic of China\\
$^{2}$Department of Physics, Hunan Normal University, \\
Changsha, 410081, People's Republic of China}
\date{}
\maketitle

\begin{abstract}
We have experimentally implemented remote state preparation (RSP) of a qubit
from a hydrogen to a carbon nucleus in molecules of carbon-13 labeled
chloroform $^{13}$CHCl$_{3}$ over interatomic distances using liquid-state
nuclear magnetic resonance (NMR) technique. Full RSP of a special ensemble
of qubits, i.e., a qubit chosen from equatorial and polar great circles on a
Bloch sphere with Pati's scheme, was achieved with one cbit communication.
Such a RSP scheme can be generalized to prepare a large number of qubit
states and may be used in other quantum information processing and quantum
computing.
\end{abstract}


\section{Introduction}

Quantum states are the carrier of quantum information. Preparing,
controlling, accessing and transmitting quantum states constitute the
important part of quantum information theory. However, the non-clone theorem%
\cite{noncloneW} of an unknown quantum state holds back the effective
utilization of quantum information. One of the striking discoveries in this
area is quantum teleportation\cite{teleport}, which transmits an unknown
quantum state from the sender (Alice) to the receiver (Bob) by a maximally
entangled channel (ebit) and two classical bit (cbit) communication, namely
1 ebit+2 cbit teleports 1 qubit. Recently, some interesting works for remote
state preparation (RSP) \cite{remoArun,remoHoi,remoBennett}, similar to
teleportation, have been theoretically proposed by Lo\cite{remoHoi}, Pati%
\cite{remoArun} and Bennett et al.\cite{remoBennett}, which can remotely
prepare a quantum state from Alice to Bob with less classical bit
communication than teleportation. In their papers, they showed that for a
special ensemble of states (e.g., qubit states chosen from the equator or
the polar of the Bloch sphere) RSP\ requires only one bit of classical
communication per qubit, exactly half that of teleportation, i.e., 1 ebit +
1 cbit transmits 1 qubit. But hitherto, though quantum teleportation has
been implemented experimentally\cite{teleportNMR,telExp1,telExp2}, RSP has
not been experimentally tested yet.

In this paper, we will experimentally demonstrate RSP for a special quantum
state (i.e., a qubit chosen from equator or polar great circles of a Bloch
sphere), with a 2-qubit sample, carbon-13 labeled chloroform $^{13}$CHCl$%
_{3},$ and complete the task that 1 ebit + 1 cbit transmits 1 qubit. For the
general case, Lo\cite{remoHoi} conjectured that the classical communication
cost of RSP is equal to that of teleportation, and their procedure would be
similar. Bennett et al.\cite{remoBennett} showed that, in the presence of a
large amount of prior entanglement, the asymptotic classical communication
cost of RSP for general states is one bit per qubit. However, the
experimental verification of the asymptotic character requires preparation
of a large amount of entanglement source, which would be difficult to
implement by the existing techniques.

\section{Remote State Preparation of a Special Ensemble and the NMR
Realization}

Consider an arbitrary single qubit state 
\begin{equation}
|\psi \rangle =\alpha |0\rangle +\beta |1\rangle ,
\end{equation}
where $\alpha =\cos (\theta /2)$ is chosen to be real and $\beta =\sin
(\theta /2)\exp (i\phi )$, a complex number. The state can be represented by
a point in a Bloch sphere with two real parameters $\theta $ and $\phi $,
and obtained by a transformation $R$ on $|0\rangle $ with 
\begin{equation}
R=\left( 
\begin{array}{cc}
\cos (\theta /2) & -\sin (\theta /2)e^{-i\phi } \\ 
\sin (\theta /2)e^{i\phi } & \cos (\theta /2)
\end{array}
\right)
\end{equation}
The task of RSP is that Alice helps Bob in a distant laboratory to prepare
the state $|\psi \rangle $ known to her but unknown to Bob.

Pati\cite{remoArun} proposed a procedure for RSP of a qubit chosen from
equatorial or polar line on a Bloch sphere. The schematic circuit for RSP\
is shown in Fig. 1(a).

Similar to teleportation, at first, Alice and Bob share a two-qubit maximal
entangled state, i.e. an EPR state, 
\begin{equation}
|\psi ^{-}\rangle _{AB}=\frac{1}{\sqrt{2}}(|0\rangle _{A}|1\rangle
_{B}-|1\rangle _{A}|0\rangle _{B}).
\end{equation}
$|\psi ^{-}\rangle _{AB}$ can be expanded in different orthogonal basis. We
choose the qubit orthogonal basis $\{|\psi \rangle ,|\psi _{\bot }\rangle \}$%
, 
\begin{equation}
\begin{array}{l}
|\psi ^{-}\rangle _{AB}=\frac{1}{\sqrt{2}}(|\psi \rangle _{A}|\psi _{\bot
}\rangle _{B}-|\psi _{\bot }\rangle _{A}|\psi \rangle _{B}), \\ 
=\frac{1}{\sqrt{2}}(|\psi \rangle _{A}U^{-1}|\psi \rangle _{B}-|\psi _{\bot
}\rangle _{A}|\psi \rangle _{B}).
\end{array}
\end{equation}
where $|\psi _{\bot }\rangle =\alpha |1\rangle -\beta ^{*}|0\rangle ,$ $%
\beta ^{*}$ is the complex conjugate of $\beta $, and $U^{-1}$ is the
inverse of $U$ which is a unitary operator to convert $|\psi _{\bot }\rangle 
$ into $|\psi \rangle .$ It is possible for Alice to perform a single
particle Von-Neumann measurement under the basis because of the state $|\psi
\rangle $ known to her. Conditional on Alice's two measurement outcomes $%
|\psi \rangle $ and $|\psi _{\bot }\rangle ,$ Bob's state is either $|\psi
\rangle $ or $U^{-1}|\psi \rangle .$ Thus, after Alice performs the
measurement, she sends the result (one classical bit information) to Bob,
who can then recover the desired state $|\psi \rangle $ by deciding to do
nothing $E$ or a unitary operation $U$ according to the received classical
information. For a special ensemble, $U$ can be an affirmative unitary
operator to transform $|\psi _{\bot }\rangle $ into $|\psi \rangle .$ For
example, if Alice choose to prepare a polar great circles on a Bloch sphere,
i. e., $\phi =0,|\psi \rangle =\cos (\theta /2)|0\rangle +\sin (\theta
/2)|1\rangle ,$ then $U=i\sigma _{y}.$ Alternatively, if Alice wish to help
Bob prepare an arbitrary equatorial state such as $|\psi \rangle =\frac{1}{%
\sqrt{2}}(|0\rangle +e^{i\phi }|1\rangle ),$ then $U=\sigma _{z}.$ However,
for a general state (both $\theta $ and $\phi $ are arbitrary), there is no
universal transformation $U$ to take $|\psi _{\bot }\rangle \rightarrow
|\psi \rangle $ by this protocol. Therefore, RSP of an arbitrary state can
be made with the success probability of 50\%.

The network of RSP\ for a special ensemble shown in Fig. 1(b) corresponds to
the schematic circuit in Fig.1(a), including three steps.

(1) {\bf Preparation of the EPR pair }$|\psi ^{-}\rangle _{AB}.$ In NMR
experiments, suppose that Alice and Bob have spin $A$ and spin $B$,
respectively, consisting of a bipartite spin quantum system in a $4\times 4$
Hilbert space $H_{A}\otimes H_{B}.$ The EPR pair can be prepared from a pure
state $|0\rangle _{A}|0\rangle _{B}$ by the operations of NOT ($N_{i}$ flips
the sign of spin $i$), the Hadamard gate ($H_{i}$ transform the states
according to ($|0\rangle \rightarrow \frac{1}{\sqrt{2}}(|0\rangle +|1\rangle
),|1\rangle \rightarrow \frac{1}{\sqrt{2}}(|0\rangle -|1\rangle )$) and
controlled-NOT ($CN_{\bar{A}B}$ flips spin $B$ if and only if $A$ is $%
|0\rangle $ ) gates. The $N_{A}$ , $H_{A}$ and $CN_{\bar{A}B}$ gates were
respectively realized by the pulse sequences $X_{A}(\pi )$, $\bar{Y}_{A}(%
\frac{\pi }{2})\bar{X}_{A}(\pi )$ and $\bar{Y}_{B}(\frac{\pi }{2}%
)Z_{B}\left( \frac{\pi }{2}\right) Z_{A}\left( \frac{\pi }{2}\right)
J_{AB}(\pi )$ (pulses applied from right to left). By simplification
methodology\cite{Ernst}, the EPR pair was realized by using the pulse
sequence

\begin{equation}
\bar{Y}_{B}(\frac{\pi }{2})J_{AB}(\pi )Y_{B}\left( \frac{\pi }{2}\right) 
\bar{X}_{B}\left( \frac{\pi }{2}\right) \bar{Y}_{A}\left( \pi \right)
X_{A}\left( \frac{\pi }{2}\right) .
\end{equation}

(2) {\bf The single particle Von-Neumann measurement under the qubit basis }$%
\{|\psi \rangle ,|\psi _{\bot }\rangle \}.$ We use a two-part procedure
inspired by Brassard et al. to perform the measurement \cite{measure}. Part
one of the procedure is to rotate from the basis $\{|\psi \rangle ,|\psi
_{\bot }\rangle \}$ into the computational basis $\{|0\rangle ,|1\rangle \}$
by the transformation $R^{+}=R^{-1},$ inverse of $R$. Part two of the
procedure is to perform a projective measurement in the computational basis.

Applying $R_{A}^{+}$ on the EPR\ pair, one gets 
\begin{equation}
R_{A}^{+}|\psi ^{-}\rangle _{AB}\rightarrow \frac{1}{\sqrt{2}}(|0\rangle
_{A}|\psi _{\bot }\rangle _{B}-|1\rangle _{A}|\psi \rangle _{B}),
\end{equation}
For a polar state, the operation 
\begin{equation}
R_{A}^{+}(\theta ,\phi =0)=\left( 
\begin{array}{ll}
\cos (\theta /2) & \sin (\theta /2) \\ 
-\sin (\theta /2) & \cos (\theta /2)
\end{array}
\right) ,
\end{equation}
which can be realized by the NMR pulse sequence $\bar{Y}_{A}\left( \theta
\right) .$ For an equatorial state, the operation 
\begin{equation}
R_{A}^{+}\left( \theta =90^{\circ },\phi \right) =\frac{1}{\sqrt{2}}\left( 
\begin{array}{ll}
1 & e^{-i\phi } \\ 
-e^{i\phi } & 1
\end{array}
\right)
\end{equation}
which can be realized by the NMR pulse sequence $X_{A}\left( \theta
_{1}\right) \bar{Y}_{A}\left( \theta _{2}\right) X_{A}\left( \theta
_{1}\right) ,$ where $\theta _{1}=\tan ^{-1}\left( \sin \phi \right) ,$ and $%
\theta _{2}=2\sin ^{-1}\left( \cos \phi /\sqrt{2}\right) .$

It is possible to simulate directly the second step in NMR exploiting
magnetic gradient pulses\cite{project}. However, it can be seen from Eq. (6)
that, the final state of Alice is one of $|0\rangle $ and $|1\rangle ,$
corresponding to the two different measurement results. Owing to the weak
measurement of NMR, we can employ the conditional unitary operation $S$,
instead of the projective measurement in the computational basis and the
post-measurement operation.

(3) {\bf Conditional unitary operation }$S${\bf .} It can be verified from
Eq. (6), that when Alice measures the outcome of spin $A$ $|0\rangle $, Bob
performs the transformation $U$ on spin $B$; when $|1\rangle $, Bob does
nothing, then two measurement results lead to Bob in $|\psi \rangle .$ For a
polar state, $U=i\sigma _{y},$ the conditional unitary operation $S=i\sigma
_{y}^{B}E_{+}^{A}+E_{-}^{A},$ is implemented by the pulse sequence

\begin{equation}
\bar{Y}_{B}\left( \frac{\pi }{2}\right) X_{B}\left( \frac{\pi }{2}\right)
J_{AB}\left( \pi \right) \bar{X}_{B}\left( \frac{\pi }{2}\right) .
\end{equation}
For an equatorial state, $U=\sigma _{z},$ the conditional unitary operation $%
S=\sigma _{z}^{B}E_{+}^{A}+E_{-}^{A},$ is implemented by the pulse sequence

\begin{equation}
\bar{Y}_{A}\left( \frac{\pi }{2}\right) \bar{X}_{A}\left( \frac{\pi }{2}%
\right) Y_{A}\left( \frac{\pi }{2}\right) \bar{Y}_{B}\left( \frac{\pi }{2}%
\right) X_{B}\left( \frac{\pi }{2}\right) Y_{B}\left( \frac{\pi }{2}\right)
J_{AB}\left( \pi \right) .
\end{equation}

(4) {\bf The measurement of the final state}. After tracing out spin $A$ the
state of spin $B$ is reduced to the expected form $\left| \psi \right\rangle
,$ which indicates the success of RSP. In principle, spin $A$ could be
traced out experimentally by applying a decoupling field during observation
of spin $B$. But this approach was found to lead to unacceptable sample
heating, with resultant shifts in resonance frequencies. Thus, the tracing
out process was implemented in software by integrating the entire multiplet
of spin B after adjusting the right phase.

\section{ The NMR Experiment\ and Results}

The RSP scheme stated above was implemented by liquid-state NMR spectroscopy
with carbon-13 labeled chloroform $^{13}CHCl_{3}$ (Cambridge Isotope
Laboratories, Inc.). To perform RSP we make use of the hydrogen nucleus $%
(^{1}H)$ as the sender (Alice) and the carbon nuclei $(^{13}C)$ as the
receiver (Bob) in the experiments, transmitting the state from $^{1}H$ to $%
^{13}C$. Spectra were recorded on a BrukerARX500 spectrometer with a probe
tuned at 125.77MHz for $^{13}C$, and at 500.13MHz for the $^{1}H$ . The
spin-spin coupling constant $J$ between $^{13}C$ and $^{1}H$ is 214.95Hz.
The relaxation times were measured to be $T_{1}=4.8sec$ and $T_{2}=0.2sec$
for the proton, and $T_{1}=17.2sec$ and $T_{2}=0.35sec$ for carbon nuclei.

The pseudo-pure state was prepared in our experiment using line-selective
pulses and the gradient-pulse techniques \cite{xhpeng}. We performed two
separate sets of experiments of the RSP process (shown in Fig.2) from $^{1}H$
to $^{13}C.$ As all NMR observables are traceless and the constant item has
no effect on the NMR signal, any point on the Bloch sphere can be expressed
as the product form $I_{x}\sin \theta \cos \phi +I_{y}\sin \theta \sin \phi
+I_{z}\cos \theta $ (apart from a constant unit matrix), which corresponds
to the observable NMR signal. Using Alice's knowledge of the qubit, an
advanced operation $R^{+}$ was performed on $^{1}H$ for a given set of $%
\theta $ and $\phi $ values. In the first set, a total of 25 qubit states
chosen from a range of points on the polar line of the Bloch sphere $\left(
\phi =0\right) $ with a $\theta $ spacing of $\pi /12$ were studied. The
second set of experiments was executed for RSP\ of 17 equatorial states on
the Bloch sphere $\left( \theta =\pi /2\right) $ with a $\phi $ spacing of $%
\pi /8.$ For each state the experimental spectra of $^{13}C$ were recorded
from the $^{13}C$ nucleus, and their real and imaginary components are
plotted in Fig. 2. The experimental results clearly show the expected line,
sine and cosine modulations. It can be seen from Fig. 2, there exists an
same initial phase offset in the fitting curves, which maybe caused by the
systemic errors, such as static magnetic field and rf field inhomogeneities
and the imperfect calibrations of rf pulses, etc.. Furthermore, we
reconstructed the density matrices of the obtained states by quantum state
tomography\cite{tomography}, and the maximal relative errors $\bigtriangleup
<18\%,$ where

\begin{equation}
\bigtriangleup =\max_{all\text{ }states}\left\{ \frac{||\rho _{theory}-\rho
_{\exp t}||}{||\rho _{theory}||}\right\} ,
\end{equation}
and $\rho _{theory}$ is the expectative state of RSP, and $\rho _{\exp t}$
is the reconstructed state by the experimental data. Besides static magnetic
field and rf field inhomogeneities and the imperfect calibrations of rf
pulses, the errors in experiments arise from decoherence due to the $T_{2}$
relaxations of nuclei and the environmental effects such as temperature.
Nevertheless, it turns out from Fig. 2 and the reconstructed density
matrices that the RSP network is effective for a special ensemble.

\section{Conclusion}

In summary, we have experimentally implemented the RSP protocol by using NMR
quantum logic gates and circuits in quantum information. In our experiments,
RSP\ of a special ensemble was successfully completed by a maximally
entangled channel with one classical bit communication, exactly half that of
teleportation, namely 1 cbit $\rightarrow $ 1 qubit. The special states
distributes on the equatorial and polar circles on the Bloch sphere. The RSP
protocol can be generalized to remotely prepare a large number of general
qubit states in the high-entanglement limit at an asymptotic cost of one bit
per qubit \cite{remoBennett}. Like teleportation, RSP can be applied not
only to pure states, but also to parts of entangled states which has been
studied by Bennett et al.\cite{remoBennett}. Although the communication
happens between spins in angstrom distance, the concept and method of
quantum information transmission should be useful for quantum computation
and quantum communication.

\begin{center}
{\bf ACKNOWLEDGMENTS}
\end{center}

We thank C. P. Sun for bringing the topic of RSP to our attention and
Xiaodong Yang, Hanzheng Yuan, Xu Zhang and Guang Lu for help in the course
of experiments.

\begin{center}
{\large Figure Captions}
\end{center}

Fig. 1 Schematic protocol for RSP proposed by Pati (a) and the corresponding
network of implementing RSP (b). $H$ represents the Hadamard gate and
conditional operation on a spin being in the $\left| 1\right\rangle $ state
and the $\left| 0\right\rangle $ state are represented by a filled circle
and an empty circle, respectively. $R^{+}$ is demonstrated in the text.

Fig. 2 Experimental results for RSP of qubits chosen from (a) the polar
line, and (b) the equatorial circle, on the Bloch sphere. Data points $%
\bigcirc $ and $*$ denote the real and imaginary parts of the NMR\ signals
from $^{13}C$, respectively. The fitting expectative curves are depicted
with the solid lines.

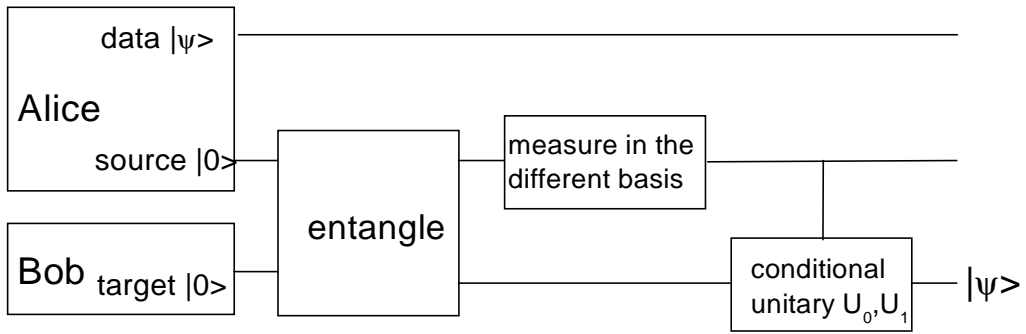
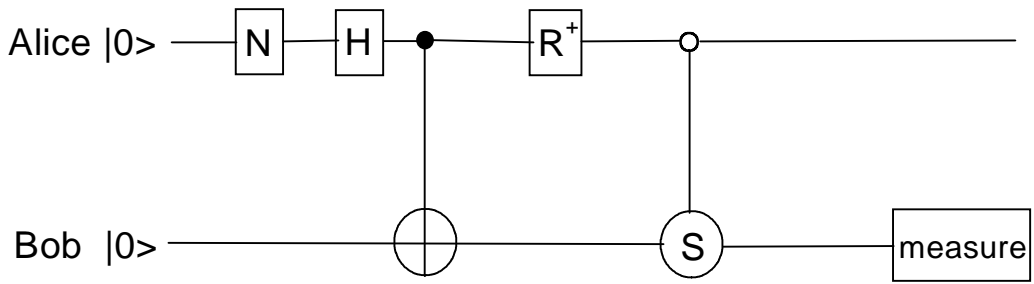
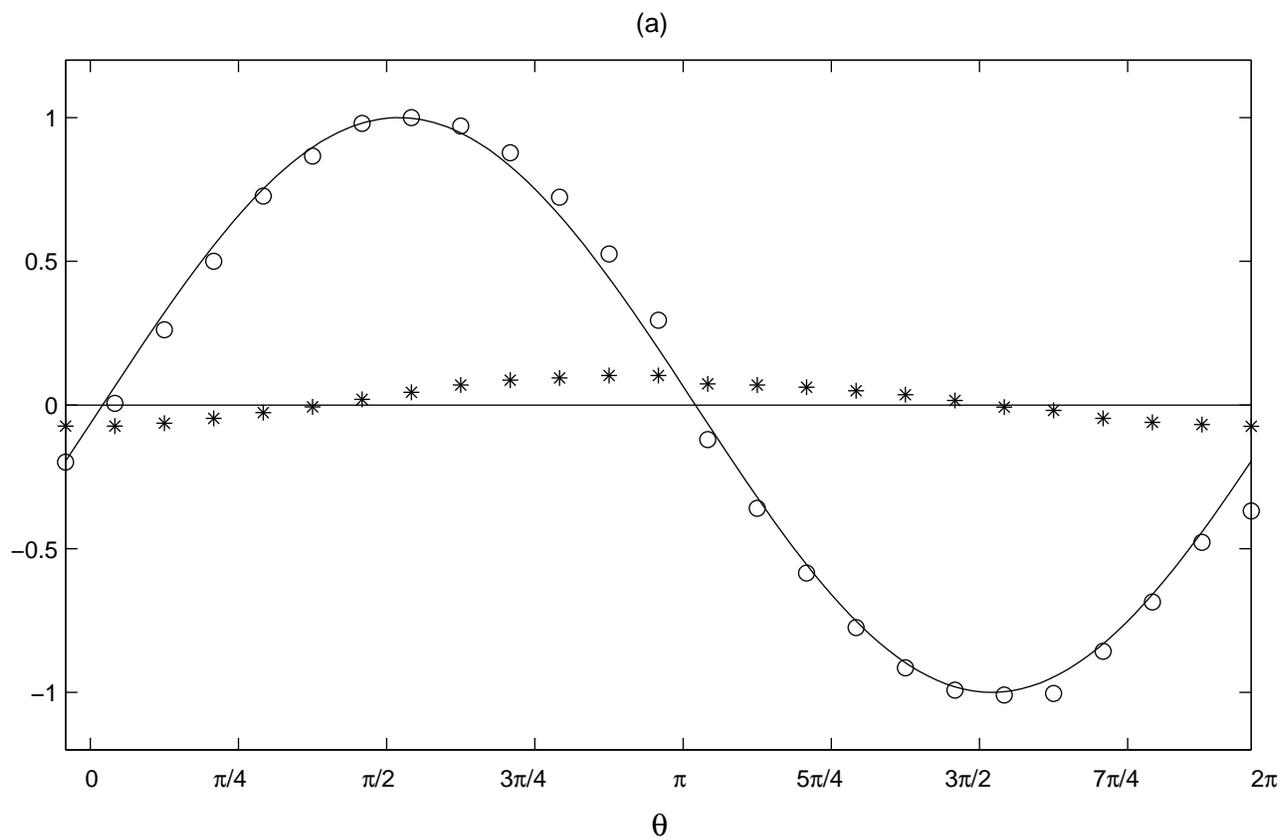
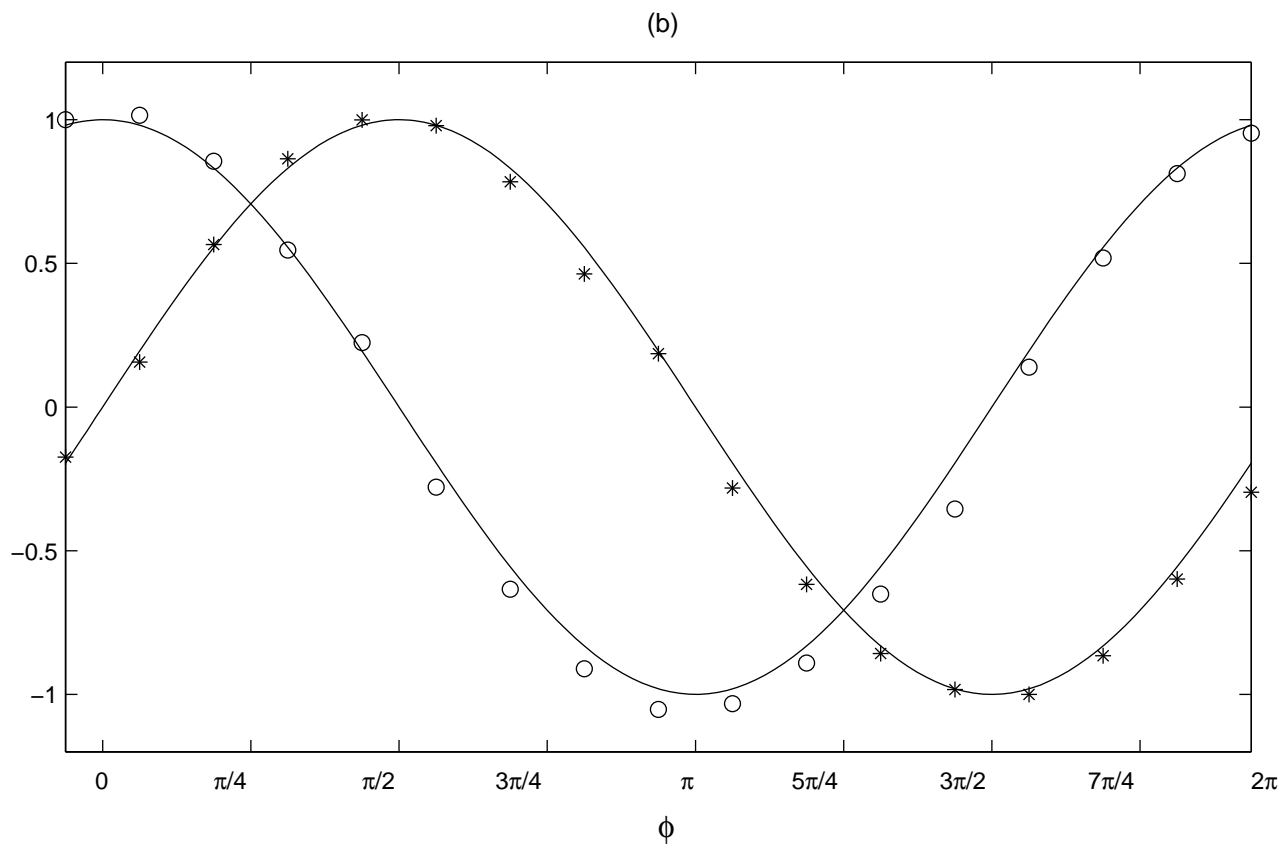
\end{document}